\documentclass[prl,twocolumn,superscriptaddress,showpacs,amsmath,amssymb]{revtex4-1}

\usepackage{graphicx}
\usepackage{latexsym}
\usepackage{amsmath}
\usepackage{amssymb}
\usepackage{amsfonts}
\usepackage{color}
\usepackage{bm}
\usepackage{verbatim}
\usepackage[]{color}
\definecolor{red}{rgb}{1,0,0}
\usepackage{framed} 
\usepackage{capt-of}
\definecolor{shadecolor}{RGB}{222,222,221}
\definecolor{MS-color}{RGB}{128,0,128}

\graphicspath{{Figs'/}}
\begin{document}

\title{Anomalous phase shift in a Josephson junction via an antiferromagnetic interlayer}

 \date{\today}

\author{D. S. Rabinovich}
\affiliation{Skolkovo Institute of Science and Technology, Skolkovo 143026, Russia}
\affiliation{Moscow Institute of Physics and Technology, Dolgoprudny, 141700 Russia}
\affiliation{Institute of Solid State Physics, Chernogolovka, Moscow
  reg., 142432 Russia}

\author{I. V. Bobkova}
\affiliation{Institute of Solid State Physics, Chernogolovka, Moscow
  reg., 142432 Russia}
\affiliation{Moscow Institute of Physics and Technology, Dolgoprudny, 141700 Russia}
\affiliation{National Research University Higher School of Economics, Moscow, 101000 Russia}

\author{A. M. Bobkov}
\affiliation{Institute of Solid State Physics, Chernogolovka, Moscow reg., 142432 Russia}

\begin{abstract}
The anomalous ground state phase shift in superconductor/antiferromagnet/superconductor Josephson junctions in the presence of the Rashba spin-orbit coupling is predicted and numerically investigated. It is found to be a consequence of the uncompensated magnetic moment at the superconductor/antiferromagnet interfaces. The anomalous phase shift exhibits a strong dependence on the value of the spin-orbit  coupling and the sublattice magnetization with the simultaneous existence of stable and metastable branches. It depends strongly on the orientation of the Neel vector with respect to the superconductor/antiferromagnet interfaces via the dependence on the orientation of the interface uncompensated magnetic moment. This effect opens a way to control the Neel vector by supercurrent in Josephson systems.
\end{abstract}

  \maketitle
 
\section{Introduction}

The current-phase relation (CPR) of a Josephson junction (JJ) in its minimal form can be written as $j=j_c \sin (\varphi-\varphi_0)$, where $j_c>0$ is the critical current of the junction and $\varphi$ is the phase difference between the superconducting leads. In the ground state, which corresponds to $j=0$, the phase difference $\varphi = \varphi_0$. Ordinary Josephson junctions are characterized by $\varphi_0 = 0$. There are various examples, including  superconductor/ferromagnet/superconductor JJs
\cite{Ryazanov2001,Buzdin2005,Oboznov2006}, JJs via nonequilibrium normal interlayer \cite{Baselmans1999}, JJs composed of d-wave superconductors \cite{VanHarlingen1995,Hilgenkamp2003}, JJs through quantum dots \cite{vanDam2006}
and gate-controlled carbon nanotubes \cite{Cleuziou2006}, where $\varphi_0=\pi$ state is realized. The so-called anomalous ground state phase shift $\varphi_0 \neq \pi n$ can also occur in systems with broken time-reversal symmetry. For more general non-sinusoidal CPR this means that condition $j(\varphi)=-j(-\varphi)$ is violated. It has been reported that such a state can be realized in Josephson junctions with ferromagnetic interlayers or under externally applied Zeeman field in the presence of a spin-orbit(SO) coupling \cite{Krive04,Asano2007,Reynoso2008,Buzdin2008,Tanaka2009,Zazunov2009,Malshukov2010,Brunetti2013,Yokoyama2014,Bergeret2015,Konschelle2015,Campagnano2015,Kuzmanovski2016,
Zyuzin2016,Silaev2017,Bobkova2017, Minutillo2018}, as well as with inhomogeneous ferromagnetic interlayers without an external SO-coupling \cite{Silaev2017,Bobkova2017,Braude2007,Grein2009,Liu2010,Kulagina2014,Moor2015,Moor2015_2,Mironov2015,Rabinovich2018,Pal2019} and for JJs between unconventional superconductors \cite{Tanaka1997,Sigrist1998}. This effect has been observed in the Josephson junctions through a quantum dot\cite{Szombati2016}, $Bi_2Se_3$ and $InAs$ interlayers \cite{Murani2017,Assouline2018,Mayer2019}. The $\varphi_0$-shift should be distinguished from the symmetric ground state phase shift $\pm|\varphi| \neq \pi n$, which does not require the time-reversal symmetry breaking and can occur due to the presence of higher harmonics in the CPR \cite{Barash1995,Yip1995}.

Furthermore, it was realized that the anomalous phase shift is of great interest for superconducting spintronics because provides a possibility for  control of magnetization by supercurrent \cite{Braude2008,Konschelle2009,Kulagina2014,Shukrinov2017,Bobkova2018,Shukrinov2018} and electrical detection of magnetization dynamics \cite{Rabinovich2019}.

Traditionally spintronic devices have been based on ferromagnetic materials. At the same time nowadays there is an emerging
subfield of spintronics research in which the central role is played by antiferromagnets \cite{Jungwirth2018}. Among them are demonstrations of experimental microelectronic memory devices \cite{Wadley2016,Olejnik2017,Bodnar2018,Meinert2018}, optical
control of antiferromagnetic spins \cite{Manz2016,Baierl2016,Bossini2016}, interplay of antiferromagnetic spintronics with topological phenomena \cite{Tang2016,Smejkal2017,Yang2017}, large anomalous Hall effect and topological Hall effect in non-collinear antiferromagnets \cite{Chen2014,Nakatsuji2015,Surgers2014}. However, the prospects of synergy between antiferromagnets and superconducting spintronics have barely been investigated. In this respect it was reported that an antiferromagnetic insulator with effectively uncompensated interface induces a spin splitting in the adjacent superconductor \cite{Kamra2018}. An enhancement of magnon-mediated superconductivity on the surface of the topological insulator was also predicted \cite{Erlandsen2019}. 

Proximity and Josephson effects through antiferromagnetic interlayers have already been investigated \cite{Demler1998,Zabel1999,Gorkov2001,Miyagawa2002,Andersen2002,Bobkova2005_1, Bobkova2005_2, Bell2003,Andersen2006,Andersen2008,Weides2009,Zaitsev2009,Zaitsev2010,Enoksen2013,Bakurskiy2015}. In particular, it was found that the Josephson current is highly sensitive to the junction length revealing $0-\pi$-transitions depending on the number of atomic layers. However, reports of the anomalous ground state phase shift in Josephson junctions through antiferromagnets are lacking. In the present paper we predict an anomalous ground state phase shift in a Josephson junction through an antiferromagnet in the presence of Rashba SO-coupling.  As it was already mentioned above anomalous phase shifts are now considered as key elements in superconducting spintronics. Therefore, our results provide an avenue to synergy between antiferromagnetic and superconducting branches of spintronics. In particular, they open a way to Josephson detection of antiferromagnetic order dynamics and its control by supercurrent.  

We have found that the anomalous phase shift only exists if the antiferromagnet is uncompensated. However, we demonstrate that it cannot be explained only by an influence of uncompensated magnetic moment,  and the antiferromagnetic order by itself is of crucial importance for obtaining essential values of the anomalous phase shift $\varphi_0$. The presence of surface magnetization, stemming from broken translational symmetry at interfaces has already been observed experimentally \cite{Takano1997,Antel1999,Hase2001,Ohldag2001,Ohldag2003}. The effect can also be relevant to 2D antiferromagnets discovered recently \cite{Gong2019,Taniguchi2018,Hu2016}, especially taking into account the large SO-coupling predicted for these materials. 

\begin{figure}[!tbh]
 \begin{minipage}[b]{\linewidth}
   \centerline{\includegraphics[clip=true,width=3.0in]{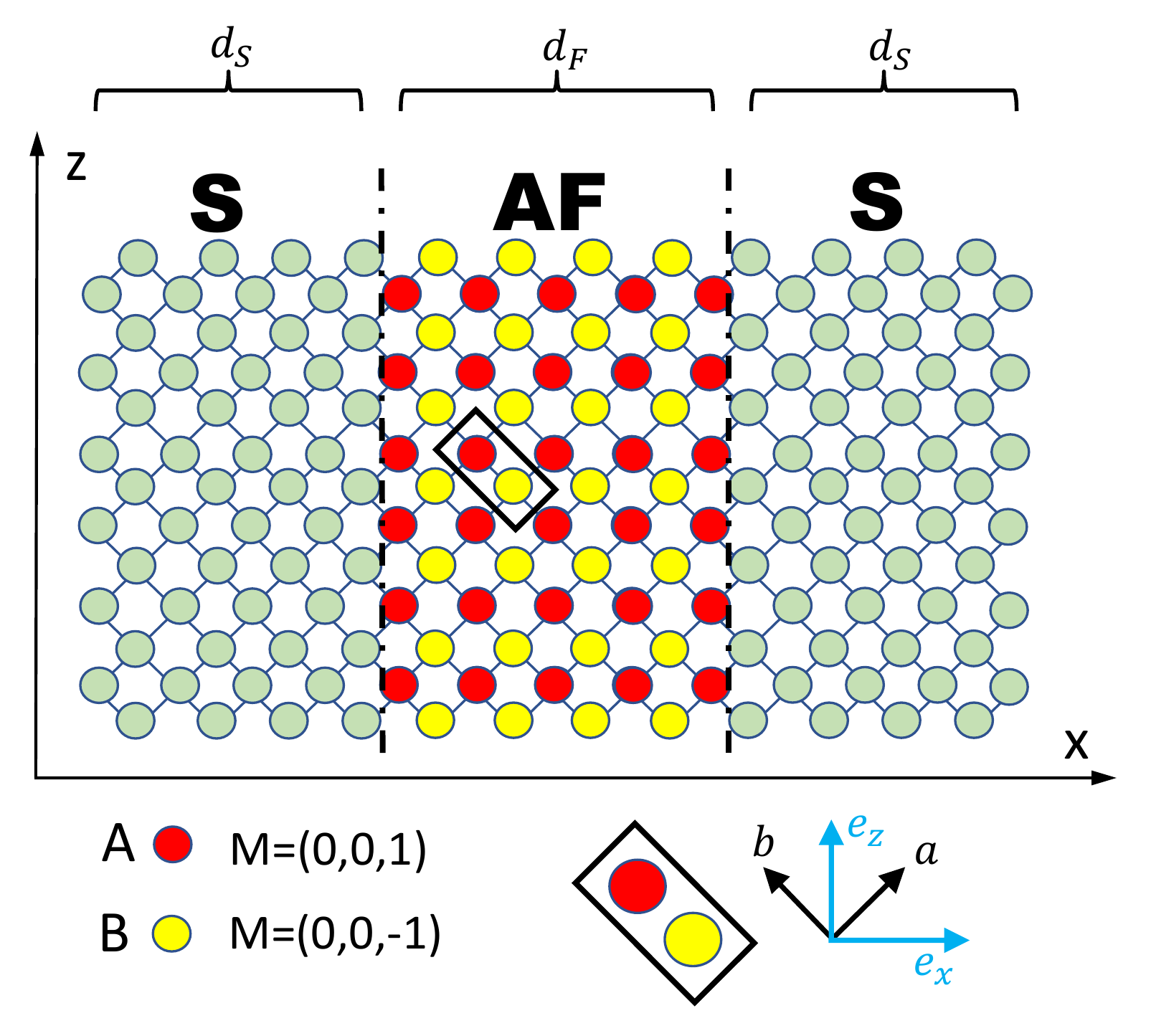}}
   \end{minipage}
      \caption{System under consideration: S/AF/S junction in (110) orientation. The presented junction is called $AA$-junction in the text because the both interface planes of the antiferromagnet are occupied by $A$-atoms. The unit cell containing two neighboring atoms belonging to different sublattices is shown by the black rectangle. The basis vectors $\bm e_{x,z}$ are shown together with the unit vectors $\bm a$ and $\bm b$ along the crystal axes.}
 \label{sketch}
 \end{figure}
 
\section{Model and method}

The sketch of the system under consideration is presented in Fig.~\ref{sketch}. We focus on junctions with (110) oriented interfaces. Different situations were studied: the junction with an uncompensated magnetic moment, shown in the sketch (this example is called by $AA$-junction) and the junction with fully compensated magnetic moment ($AB$-junction). 
To study the superconductor/antiferromagnet/superconductor (S/AF/S) Josephson junction we
consider the following two-dimensional  mean-field Hamiltonian on a square lattice

\begin{eqnarray}
H= - t \sum \limits_{\langle ij \rangle \sigma}  c_{i\sigma}^\dagger c_{j\sigma} + \sum \limits_{\langle ij \rangle } (\Delta_{i} c_{i\uparrow}^\dagger  c_{j\downarrow}^\dagger + H.c.) - \nonumber \\
\mu \sum \limits_{i \sigma} \hat n_{i\sigma} + \sum \limits_i \hat c_{i}^\dagger \bm m_{i} \bm \sigma \hat c_{j} + H_R,~~~~~~
\label{hamiltonian}
\end{eqnarray}
where $\Delta_{i}$ and $\bm m_i$ denote the superconducting and magnetic
order parameters, respectively. For a square lattice with the crystal
coordinate axes $a$ and $b$, the staggered magnetic order parameter takes the form $\bm m_i = (-1)^{i_a+i_b}\bm m$. Here $c_{i\sigma}^\dagger$ creates an electron of spin $\sigma$ on the site $i$, $t$ denotes the nearest-neighbor hopping integral, $\mu$ is the filling factor, and $\hat n_{i\sigma} =   c_{i\sigma}^\dagger  c_{i\sigma}$ is the particle number operator on site $i$. $\hat c_i = (c_{i \uparrow},c_{i \downarrow})^T$ and $\bm \sigma = (\sigma_x,\sigma_y, \sigma_z)$ is a vector of Pauli matrices in spin space.    

Further, $H_R$ is the Rashba SO coupling due to the broken structural inversion symmetry and is relevant to common experimental geometries in which a thin AF film is on the substrate:
\begin{eqnarray}
H_R = -i \frac{V_{R}}{\sqrt 2} \sum \limits_i \Bigl[ \hat c_{i}^\dagger (\sigma_x + \sigma_z) \hat c_{i+\bm b}+ \nonumber \\
\hat c_{i}^\dagger  (\sigma_x - \sigma_z) \hat c_{i+\bm a} - H.c. \Bigr],
\label{rashba_term}
\end{eqnarray}
where $\bm a$ and $\bm b$ are unit vectors along the corresponding crystal coordinate axes. 

We diagonalize Hamiltonian (\ref{hamiltonian})  with the Bogoliubov transformations $c_{i \sigma}=\sum\limits_n u^i_{n\sigma}\hat b_n+v^{i*}_{n\sigma}\hat b_n^\dagger$, where sum is taken over positive eigenstates of (\ref{hamiltonian}). To simplify the problem we consider the junction to be infinite along the interface. We take into account the magnetic crystal symmetry by introducing a unit cell which contains two neighboring atoms belonging to different magnetic sublattices A and B  (see Fig  \ref{sketch}). The Fourier transformation is taken to be of the form:

\begin{eqnarray}
u_{n,\sigma}^{\bm j, A(B)}=\dfrac{a}{\sqrt{2}\pi}\int\limits_{-\pi/\sqrt{2}a}^{\pi/\sqrt{2}a} dk_z e^{ik_z a\sqrt{2}(j_z \pm 1/4)} u_{n,\sigma}^{ j_x, A(B)}(k_z).~~~~
\end{eqnarray}
The same expression is also valid for $v_{n,\sigma}^{\bm j, A(B)}$. The vector $\bm j = (j_x,j_z)$ denotes coordinates of the cells, $j_{x,z}$ are measured in units of the corresponding basis vectors $\bm e_{x,z}$, see Fig.~\ref{sketch}. For definiteness we identify cell positions with positions of the associated site A.  

The corresponding Bogoliubov-de Gennes(BdG) equations take the form 

\begin{align}
 -\mu \hat u^{j_x, A(B)}_{n}-2t\cos\frac{k_z}{\sqrt{2}}(\hat u^{j_x, B(A)}_{n}+\hat u^{j_x\mp1, B(A)}_{n})+\nonumber\\
\bm m_{j_x}^{A(B)} \bm \sigma \hat u^{j_x, A(B)}_{n}+\Delta_{j_x}^{A(B)}i \sigma_y \hat v^{j_x, A(B)}_{n}+\dfrac{V_R}{\sqrt{2}} \times\nonumber\\
2\sin\frac{k_z}{\sqrt{2}} \sigma_x (\hat u^{j_x, B(A)}_{n}+\hat u^{j_x\mp1, B(A)}_{n})\pm i \dfrac{V_R}{\sqrt{2}} \times \nonumber\\
2\cos\frac{k_z}{\sqrt{2}}\sigma_z(\hat u^{j_x, B(A)}_{n}-\hat u^{j_x\mp 1, B(A)}_{n})=\varepsilon_n \hat u^{j_x, A(B)}_{n},
\label{BdG}
\end{align}

where $\hat u(\hat v)_n^{j_x,A(B)} = (u(v)_{n.\uparrow}^{j_x,A(B)}, u(v)_{n,\downarrow}^{j_x,A(B)})^T$. The second equation is obtained by the transformation $\hat u^{j_x,A(B)}_{n} \leftrightarrow \hat v^{j_x,A(B)}_{n}$,   $\varepsilon_n \to -\varepsilon_n$, $V_R \to -V_R$, $\Delta^{A(B)}_{j_x}\to\Delta^{A(B)*}_{j_x}$ and $\bm \sigma \to \bm \sigma^T$.
The electric current from A atom in $j_x$ unit cell is 
\begin{eqnarray}
J_x = \dfrac{2a e}{\pi}{\rm Re}\sum \limits_n \int\limits_{-\pi/\sqrt{2}a}^{\pi/\sqrt{2}a} dk_z \Big\{f(\varepsilon_n) \hat u^{j_x,B\dagger}_{n}(it_{k_z}-  \nonumber \\
\hat V_R)\hat u^{j_x,A}_{n}-
(1-f(\varepsilon_n))\hat v^{j_x,B\dagger}_{n}(it_{k_z}+\hat V_R)\hat v^{j_x,A}_{n}\Bigl\},~~~~~~~~~~~
\label{current}
\end{eqnarray}
where $t_{k_z} = t \cos(k_z a/\sqrt 2)$ and we define the SO-operator in the spin space as $\hat V_R = (V_R/\sqrt 2)[\sigma_z \cos(k_z a/\sqrt 2)+\sigma_x i \sin (k_z a/\sqrt 2)]$, and $f(\varepsilon_n)$ is the Fermi distribution function. For numerical calculations we assume that temperature is zero.
\begin{figure}[!tbh]
 \begin{minipage}[b]{\linewidth}
   \centerline{\includegraphics[clip=true,width=2.8in]{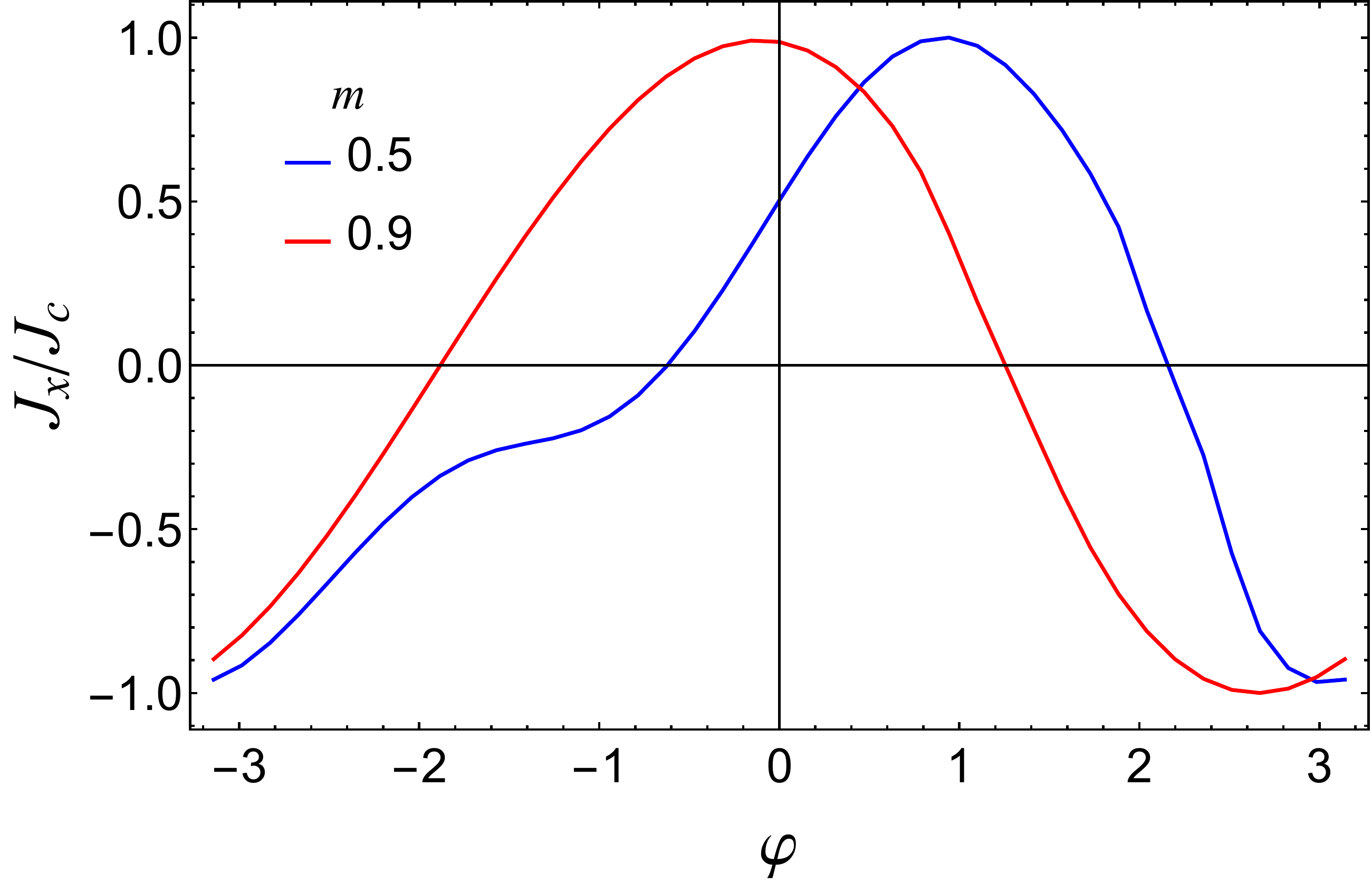}}
   \end{minipage}
      \caption{CPRs for AA-junction. The current is normalized to its critical value. Other parameters are $V_R=0.28$, $d_F=21$, where $d_F$ is the number of vertical atomic layers in the interlayer. Here and below the number of vertical atomic layers in the superconductors $d_S = 40$, $\mu = 0.3 t$, $\Delta = 0.1t$. $m$ and $V_R$ are measured in units of $t$.}
 \label{CPR}
 \end{figure}
 
\section{Results}

At first the spin quantization axis is chosen along the S/AF interface and the magnetic moment of A-sites is directed along this axis, while the moment of B-sites is the opposite. The dependence of $\varphi_0$ on the orientation of the Neel vector is discussed below. A typical example of CPR calculated numerically for the $AA$-junction according to Eq.~(\ref{current}) is presented in Fig.~\ref{CPR} for two different values of $m$. The CPR is close to a shifted sinusoidal shape for large values of $m$, while the higher harmonics are more pronounced for smaller $m$ values. For the $BB$-junction corresponding to the same parameters $j^{BB}(\varphi) = -j^{AA}(-\varphi)$, therefore the value of the anomalous phase shift is just the opposite. The $AB$-junction exhibits no anomalous phase shift. Therefore, we conclude that the anomalous phase shift is a consequence of the presence of an uncompensated magnetic moment. The (100) oriented interface has also been investigated, however we have found zero anomalous ground state phase in this case. This also indicates the necessity of the uncompensated magnetic moment, which is absent for this orientation.  Moreover, our results demonstrate that in the presence of the uncompensated interface magnetic moment the Josephson energies are different for opposite directions of the Neel vector (see below), while these states are degenerate in the absence of the uncompensated moment, for example, for the $AB$-junction. 

Physically the anomalous phase shift is a manifestation of the inverse magnetoelectric (or spin-galvanic) effect, typical for Josephson junctions \cite{Konschelle2015}. In superconducting systems the magnetoelectric effects are allowed in the presence of nonzero  Lifshitz-type term in the free energy of the system. At small values of the averaged exchange field $\bm h$ it takes the form $F_L = h^a \chi_i^a \partial \varphi / \partial x_i $  and couples the exchange field to the spatial derivatives of the superconducting phase $\partial \varphi / \partial x_i$ via the SO tensor $\chi_i^a$. The presence of an uncompensated magnetic moment in our Josephson junction  allows for nonzero average value of the Lifshitz-type term.

\begin{figure}[!tbh]
 \begin{minipage}[b]{\linewidth}
   \centerline{\includegraphics[clip=true,width=3.0in]{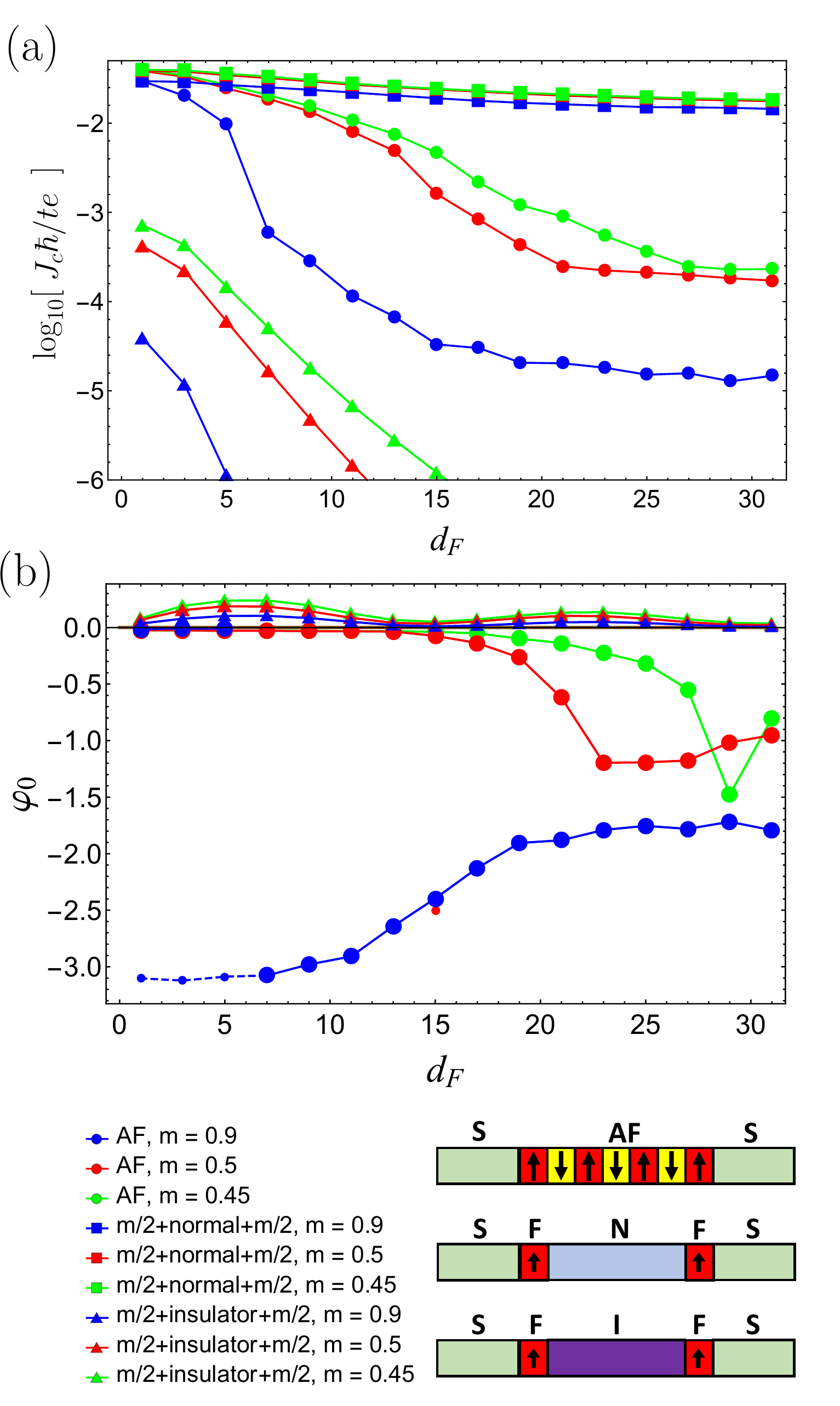}}
   \end{minipage}
      \caption{(a) Critical Josephson current and (b) ground state phase shift $\varphi_0$ as functions of $d_F$ for the AA-type S/AF/S junction with different $m$ in comparison to the critical current for "m/2+normal+m/2" and "m/2+insulator+m/2" interlayers having the same total magnetic moments. Here $V_R=0.28$.  In panel (b) stable (metastable) branches of $\varphi_0$ are plotted by solid (dashed) lines and by big (small) dots. The ground state phase shift $\varphi_0$ for "m/2+normal+m/2" model cannot be distinguished from zero in panel (b), therefore it is not shown by any symbols.{\color{red}}}
 \label{comparison}
 \end{figure}

Naively, one can suppose  that only the value of the uncompensated magnetic moment matters for appearance of the anomalous phase shift. Then the antiferromagnetic order by itself is not important and the staggered magnetic moment can be effectively replaced by its zero averaged value. Following this consideration we compare our results for the antiferromagnetic interlayer to the results obtained in the framework of two models. The first one is where  uncompensated magnetic moment is distributed symmetrically between the both interfaces with superconductors and the inner part of the interlayer is made of a normal metal, as it is shown schematically  in the bottom part of  Fig.~\ref{comparison}.
This model results from a direct averaging of the antiferromagnet magnetization, as if the interlayer was a normal metal, and we call it by "m/2+normal+m/2". We have also considered different variations of this model such as a "normal+m" model, where the uncompensated magnetic moment is not distributed symmetrically, but is placed at one of the interfaces. The results are the same. However, for the parameters under consideration the antiferromagnetic interlayer is in the insulating regime and, therefore, the model with a normal interlayer is probably not quite appropriate. For this reason in the framework of the second model the uncompensated magnetic moment is again distributed symmetrically between the interfaces, but the normal inner part of the interlayer is replaced by the insulating layer with the same gap in the electron spectrum as in the antiferromagnet for the given parameters. It is called the "m/2+insulator+m/2" model.

In Fig.~\ref{comparison} we compare the values of the critical Josephson current and the anomalous ground state phase shifts for the antiferromagnetic interlayer and for these two models. The total widths of the interlayers are the same. It can be seen that the critical current of the S/AF/S junction is strongly suppressed with respect to the m/2+normal+m/2 system. It is natural because the antiferromagnet is in the insulating regime. At the same time the critical current is much  higher than for the "m/2+insulator+m/2" model.  

In the framework of the present study it is more important to compare values of the anomalous phase shifts, which result from all these models. The results of this comparison are shown in Fig.~\ref{comparison}(b). It can be seen that none of the model systems can describe adequately the behavior of the antiferromagnet. The anomalous phase shift is nonzero as for the antiferromagnetic interlayer, so as for the both model systems considered. However, it is negligibly small for the "m/2+normal+m/2" model. It is also quite small for the "m/2+insulator+m/2" model and does not reach large intermediate values between $0$ and $\pi$. These results can be understood from the simple qualitative picture discussed below. 

For simplicity let us consider the simplest sinusoidal form of the CPR: $j=j_c \sin (\varphi - \varphi_0)$, which can be rewritten as $j=j_o \sin \varphi + j_a \cos \varphi$ via the conventional $j_o$ and anomalous $j_a$ Josephson currents, which are connected to the critical current and the anomalous phase shift as $j_c^2 = j_o^2 + j_a^2$ and $\tan \varphi_0 = -j_a/j_o$.  Due to the small value of the total uncompensated magnetic moment $j_a$ is typically rather small as for antiferromagnetic case, so as for the model systems. This is in contrast to ferromagnetic systems, where the total magnetic moment and, consequently, the strength of magnetoelectric effects, which is measured by the ratio $j_a/j_o$, can be directly controlled by the interlayer length. Here we have no such possibility. At the same time the behavior of the conventional Josephson current $j_0$ is qualitatively close to the case of zero SO coupling, where $j_o$ can be very small in the vicinity of $0$-$\pi$ transitions \cite{Andersen2006,Enoksen2013}. In this case $\varphi_0$ can have arbitrary values even if $j_a$ is small, as it is demonstrated in Fig.~\ref{comparison}(b). 

 \begin{figure}[!tbh]
 \begin{minipage}[b]{\linewidth}
   \centerline{\includegraphics[clip=true,width=2.8in]{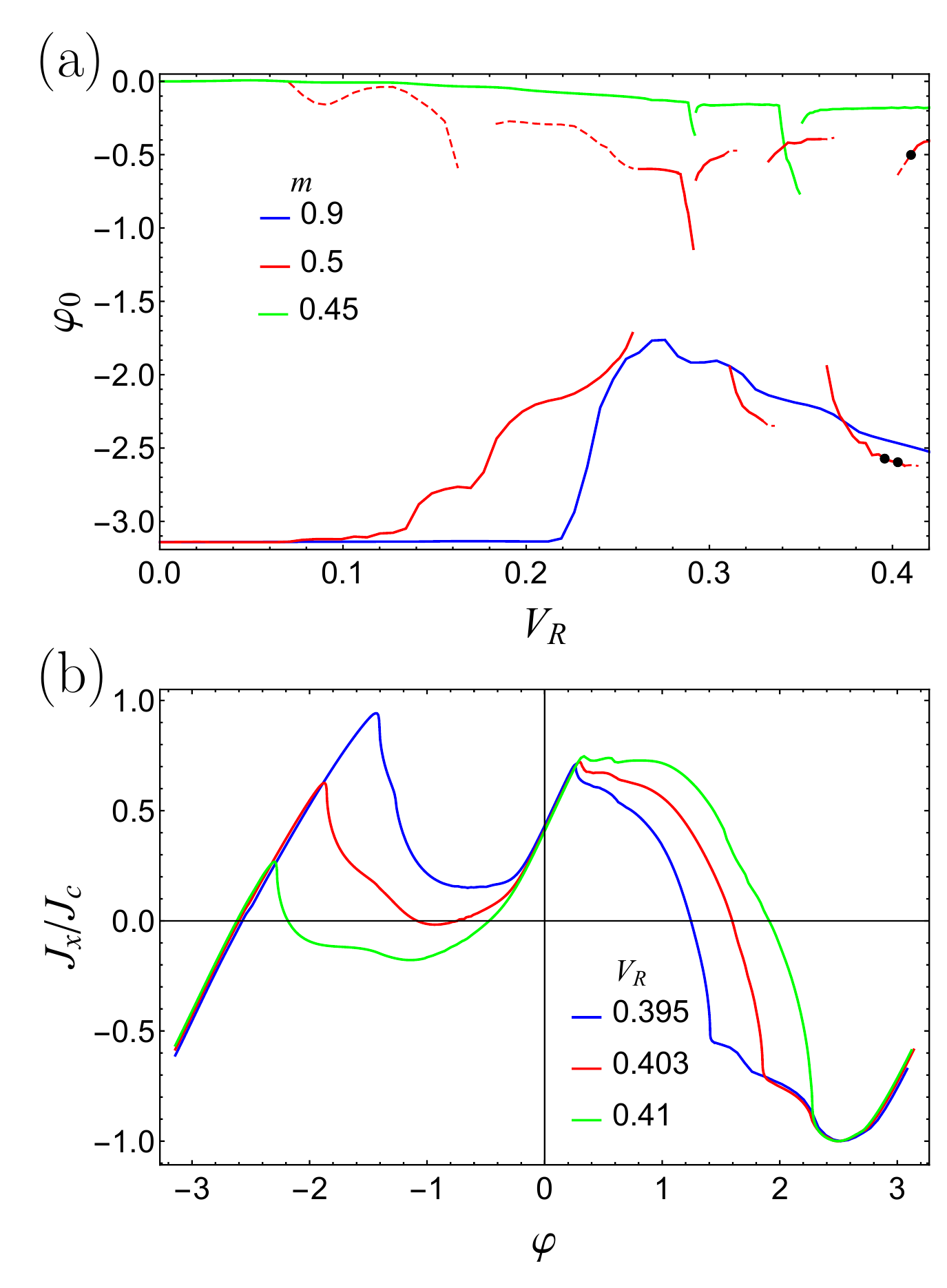}}
   \end{minipage}

      \caption{(a) $\varphi_0$ as a function of $V_R$ for different $m$. $d_F=21$. Stable (metastable) branches of $\varphi_0$ are plotted by solid (dashed) lines. (b) CPRs taken at $V_R$ marked by black points in panel (a).}
 \label{VR}
 \end{figure}
  
  Motivated by the fact that the value of the Rashba coupling can be controlled by electrical gating, we investigate the dependence of the anomalous ground state phase shift on the value of the Rashba SO-coupling. It is presented in Fig.~\ref{VR}(a). First of all, it is demonstrated that $\varphi_0$ strongly depends on $V_R$ and varies in a wide range of values between $-\pi$ and $0$. The most interesting feature is that in a particular range of the on-site magnetization values we observe simultaneous existence of several $\varphi_0$ branches. The stable branches, corresponding to the global minima of the Josephson energy  $2eE_J=\hbar \int J_x(\varphi) d\varphi$, are plotted by solid lines, while metastable solutions, corresponding to local minima, are shown by dashed lines. 
  
  To demonstrate the transition between the branches in details, in Fig.~\ref{VR}(b) we plot CPRs for three close values of $V_R$, marked by black points in Fig.~\ref{VR}(a). It can be seen that in this range of $V_R$ a new pair of zero crossing at the CPR occurs. One of them corresponds to the local minimum and the other one to the maximum of the Josephson energy. Upon further increase of $V_R$ the local maximum becomes global one, what corresponds to the abrupt change of the stable branch.  Therefore, it can be deduced that the multi-valued behavior of the anomalous phase shift as a function of the SO coupling is a consequence of the strongly non-sinusoidal behavior of CPRs, which practically always takes place in the parameter regions of $0$-$\pi$ transitions, where the most pronounced values of the anomalous phase shift should be expected, as it is discussed above.
  
 \begin{figure}[!tbh]
 \begin{minipage}[b]{\linewidth}
   \centerline{\includegraphics[clip=true,width=2.8in]{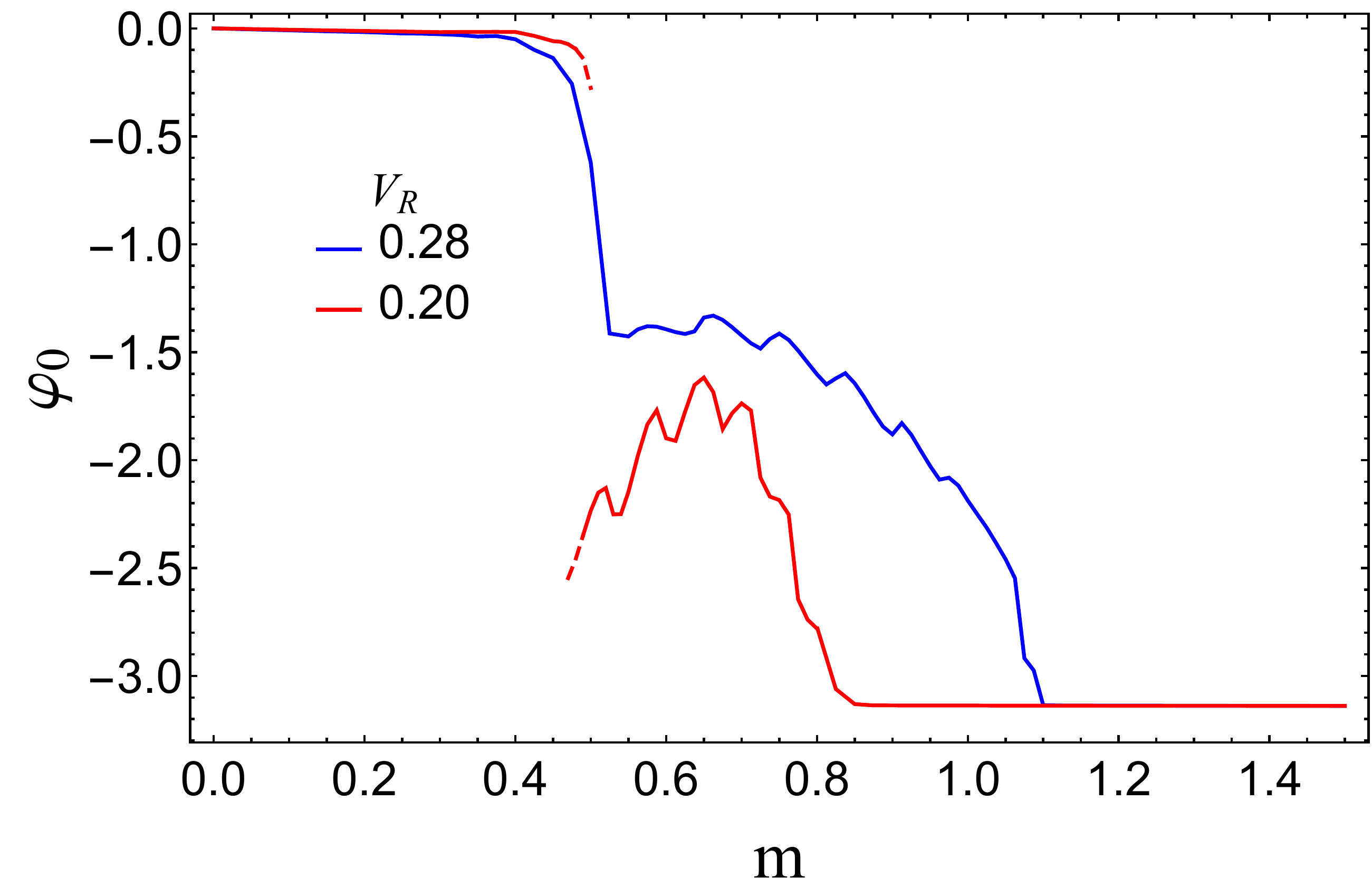}}
   \end{minipage}
        \caption{$\varphi_0$ as a function of $m$ for two different values of $V_R$. $d_F=21$. Stable (metastable) branches of $\varphi_0$ are plotted by solid (dashed) lines.}
 \label{anomalous_m}
 \end{figure} 
 
The dependence of the anomalous phase shift on the value of on-site magnetization $m$ is presented in Fig.~\ref{anomalous_m}. It is clearly seen that the system evolves between $0$ and $\pi$ states as a function of $m$ and the transition occurs via a wide region of intermediate $\varphi_0$-states. It can also be seen that $\varphi_0$ is a highly nonlinear function of $m$, what is in sharp contrast to the available theoretical and experimental results on the anomalous phase shift in Josephson junctions with low-dimensional ferromagnetic interlayers or under the applied in-plane magnetic field in the presence of the Rashba SO-coupling \cite{Buzdin2008,Konschelle2015,Assouline2018,Mayer2019}.  It is also a direct consequence of the fact that for antiferromagnets the strength of the magnetoelectric effect is determined not by the value of the sublattice magnetization $m$, but by the uncompensated magnetic moment, which is rather small and can lead to large values of the anomalous phase only when the system is close to a $0$-$\pi$ transition. Here we also observe multi-valued behavior of the anomalous phase shift with stable and metastable branches.

 \begin{figure}[!tbh]
 \begin{minipage}[b]{\linewidth}
   \centerline{\includegraphics[clip=true,width=2.8in]{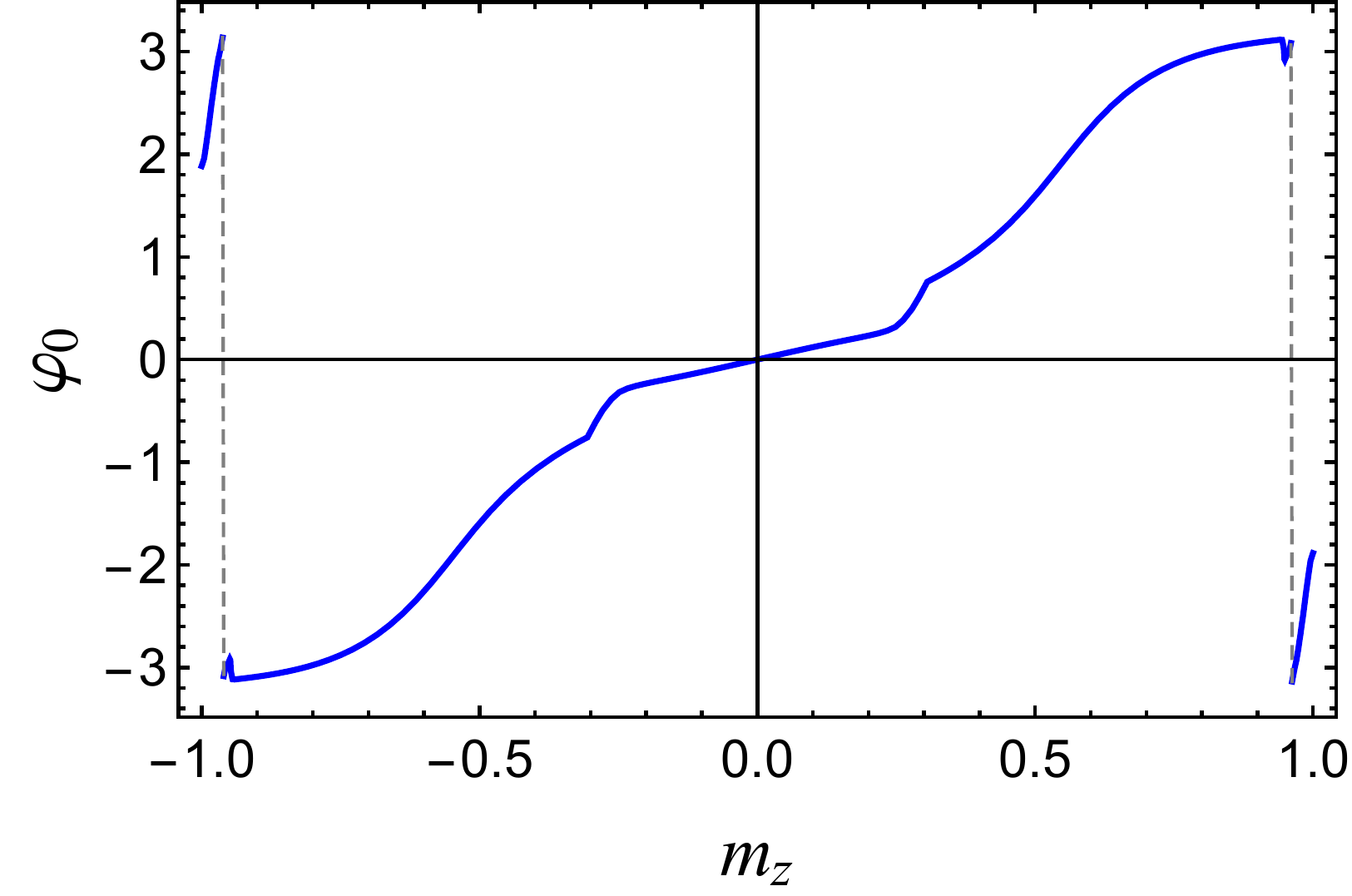}}
   \end{minipage}
        \caption{$\varphi_0$ as a function of $m_z$  component of the $A$-site  magnetization for $d_F=21$, $m=0.9$ and $V_R=0.28$.}
 \label{orientation}
 \end{figure}

Up to now we have discussed the anomalous phase shift obtained for the case when the Neel vector of the antiferromagnet is directed along the S/AF interface, that is perpendicular to the Josephson current direction. In Fig.~\ref{orientation} we demonstrate that the anomalous phase shift manifests strong dependence on the angle $\alpha$ between the $A$-site magnetization $\bm m$ and the interface (or, in other words, between the Neel vector and the interface). The Neel vector rotates in the $(x,z)$-plane. When the component of the Neel vector along the interface vanishes, $\varphi_0 = 0$.  The nature of the dependence of the anomalous phase shift on the Neel vector orientation is dictated by the Lifshitz-type term: the symmetry of the SO tensor $\chi_i^a$ is determined by the symmetry of the underlying SO coupling. In the considered case of Rashba SO coupling the only nonzero elements of $\chi_i^a$ are $\chi_x^z = -\chi_z^x$. As a result, the anomalous phase shift, which is a phase difference along the x-direction can only be coupled to $m_z$. Therefore, the anomalous phase shift is proportional to $m_z$, at least at small $m_z$, where the linear approximation is suitable. In principle, this dependence of the anomalous phase shift on the Neel vector orientation opens a different direction for studying prospects of the Neel vector control by supercurrent.

\section{conclusion}
We have predicted and numerically investigated the anomalous ground state phase shift in S/AF/S Josephson junctions in the presence of the Rashba SO-coupling. It is found that the nonzero anomalous phase shift takes place only in the presence of the uncompensated magnetic moment at S/AF interfaces. Large values of the anomalous phase shift are the consequence of the critical current suppression in the vicinity of $0$-$\pi$ transitions in the S/AF/S junction. The anomalous phase shift exhibits strong dependence on the value of the SO-coupling and the sublattice magnetization with simultaneous existence of stable and metastable branches. One of the most interesting findings is a strong dependence of the anomalous phase shift  and, therefore, Josephson energy on the orientation of the Neel vector with respect to S/AF interfaces.  In particular, the degeneracy of the Josephson energy for opposite directions of the Neel vector is removed. The interface's uncompensated magnetic moment does not require any cost in the antiferromagnetic exchange energy, as opposed to the uncompensated moment in the bulk of the antiferromagnet. Due to this fact our finding  opens a way to the Neel vector control by supercurrent in Josephson systems. By analogy with ferromagnetic systems \cite{Rabinovich2019} the presence of the anomalous phase shift provides a possibility of electrical detection of the Neel vector dynamics.  

\section {Acknowledgements}
The research of I.V.B and A.M.B has been carried out within the state task of ISSP RAS with the support by RFBR grant 19-02-00466. D.S.R acknowledges the support by RFBR grants 19-02-00466 and 19-02-00898. I.V.B. and D.S.R. also acknowledge the financial support by Foundation for the Advancement of Theoretical Physics and Mathematics “BASIS”.

\end{document}